\documentclass{aa}

\usepackage{graphicx}
\usepackage{supertabular}
\usepackage{epsfig}
\usepackage{natbib}
\usepackage[varg]{txfonts}
\usepackage{hyperref}
\usepackage[normalem]{ulem}
\usepackage{multirow}
\usepackage{longtable}
\usepackage{pdflscape}
\usepackage{color,soul}
\usepackage{amssymb}

\definecolor{MyRed}{rgb}{0.9,0.0,0.0} 
\definecolor{MyLightRed}{rgb}{1.0,0.0,0.0} 
\definecolor{MyPink}{rgb}{1.0,0.08,0.45} 
\definecolor{MyDarkBlue}{rgb}{0,0.08,0.45} 
\definecolor{MyDarkGreen}{rgb}{0,0.5,0.0} 

\newcommand{\logg}{\ensuremath{\log g}}

\newcommand{\moh}{\ensuremath{[\mathrm{M/H}]}}
\newcommand{\feoh}{\ensuremath{[\mathrm{Fe/H}]}}

\newcommand{\Teff}{\ensuremath{T_{\mathrm{eff}}}}

\newcommand{\beq}{\begin{equation}}
\newcommand{\eeq}{\end{equation}}

\newcommand{\FeoH}{\ensuremath{\left[\mathrm{Fe}/\mathrm{H}\right]}}

\newcommand{\COBOLD}{{\tt CO$^5$BOLD}}
\newcommand{\LHD}{{\tt LHD}}

\newcommand{\ATLAS}{{\tt ATLAS9}}

\newcommand{\SYNTHE}{{\tt SYNTHE}}

\newcommand{\LINFOR}{{\tt Linfor3D}}

\newcommand{\twodFH}{{\tt 2dF/HERMES}}

\begin{document}

\title{Abundance of zinc in the red giants of Galactic globular cluster 47~Tucanae}

\author{
   A.\,\v{C}erniauskas\inst{1}
   \and A.\,Ku\v{c}inskas\inst{1}
   \and J.\,Klevas\inst{1}
   \and P.\,Bonifacio\inst{2}
   \and H.-G.\,Ludwig\inst{3,2}
   \and E.\,Caffau\inst{2}
   \and M.\,Steffen\inst{4}   
   }

\institute{
       Astronomical Observatory of Vilnius University, Saul\.{e}tekio al. 3, Vilnius LT-10257, Lithuania \\
           \email{algimantas.cerniauskas@tfai.vu.lt}
           \and 
           GEPI, Observatoire de Paris, Universit\'e PSL, CNRS, 5 Place Jules Janssen, 92190 Meudon, France
           \and
           Zentrum f\"ur Astronomie der Universit\"at Heidelberg, Landessternwarte, K\"onigstuhl 12, 69117 Heidelberg, Germany       
           \and
       Leibniz-Institut f\"ur Astrophysik Potsdam, An der Sternwarte 16, D-14482 Potsdam, Germany
           }

\date{Received 18 April 2018; accepted 30 May 2018}

%============================================================================================================
\abstract
        % context heading (optional) % {} leave it empty if necessary
        {}
        % aims heading (mandatory)
        {We investigate possible relations between the abundances of zinc and the light elements sodium, magnesium, and potassium in the atmospheres ofred giant branch (RGB) stars of the Galactic globular cluster 47~Tuc and study connections between the chemical composition and dynamical properties of the cluster RGB stars.} 
        % methods  heading (mandatory)
        {The abundance of zinc was determined in 27 RGB stars of 47~Tuc using 1D~local thermal equilibrium (LTE) synthetic line profile fitting to the high-resolution \twodFH\ spectra obtained with the Anglo-Australian Telescope (AAT). Synthetic spectra used in the fitting procedure were computed with the \SYNTHE\ code and 1D~\ATLAS\ stellar model atmospheres.} 
        % results heading (mandatory)
        {The average 1D~LTE zinc-to-iron abundance ratio and its RMS variations due to star-to-star abundance spread determined in the sample of 27 RGB stars is $\langle{\rm[Zn/Fe]}\rangle^{\rm 1D~LTE}=0.11\pm0.09$. We did not detect any statistically significant relations between the abundances of zinc and those of light elements. Neither did we find any significant correlation or anticorrelation between the zinc abundance in individual stars and their projected distance from the cluster center. Finally, no statistically significant relation between the absolute radial velocities of individual stars and the abundance of zinc in their atmospheres was detected. The determined average [Zn/Fe]$^{\rm~1DLTE}$ ratio agrees well with those determined in this cluster in earlier studies and nearly coincides with that of Galactic field stars at this metallicity. All these results suggest that nucleosynthesis of zinc and light elements proceeded in separate, unrelated pathways in 47~Tuc.}
        % conclusions heading (optional),  leave it empty if necessary
        {}


\authorrunning{\v Cerniauskas et al. }
\titlerunning{Zn abundance in RGB stars in 47~Tuc}

\maketitle

%%%%%%%%%%%%%%%%%%%%%%%%%%%%%%%%%%%%%%%%%%%%%%%%%%%%%%%%%%%%%%%%%%%%%%%%%%%%%%%%%%%%%%%%%
\section{Introduction}
%%%%%%%%%%%%%%%%%%%%%%%%%%%%%%%%%%%%%%%%%%%%%%%%%%%%%%%%%%%%%%%%%%%%%%%%%%%%%%%%%%%%%%%%%

Zinc has traditionally been considered as a transition element between the Fe-peak and light $s$-process species. It is thought that zinc is produced through a variety of channels: explosive nucleosynthesis in type II supernovae (SNe~II) and type Ia~SNe (SNe Ia), hypernovae, the main and the weak components of the $s$-process \citep[see, e.g.,][]{SGC91,MRB93,MKS02,KUN06}.

Observational and theoretical studies of the Galactic evolution of zinc performed during the past decade have confirmed that zinc is synthesized through multiple channels in stars that belong to different Galactic populations. These investigations have shown that the [Zn/Fe] ratio reaches up to $\approx+0.5$ in halo stars with $\feoh<-3.0$ \citep{CDS04}. In the thin and thick disks, however, its value decreases with increasing metallicity and becomes sub-solar at $\feoh\approx0.0$, with significant scatter in [Zn/Fe] ratios at this metallicity in giant stars that most likely belong to the thin disk \citep[e.g.,][]{DCS17}. \citet{NS11} have found that solar neighborhood stars with low and high $\alpha$-element abundances were characterized with correspondingly low and high [Zn/Fe] ratios. More recently, \citet{DCS17} have shown that stars located at galactocentric distances $< 7$ \,kpc seem be to be depleted in zinc, down to ${\rm [Zn/Fe]}\approx-0.3$, which agrees with the values found in the Galactic bulge stars \citep{BFd15,dSBF18}. These and other observations suggest a related origin of zinc and $\alpha$-elements \citep[e.g.,][]{dSBF18}. However, [Zn/Fe] trends at low metallicity point to a production of zinc in high-energy core-collapse supernovae, that is, hypernovae \citep[][]{KUN06}. The significant scatter observed in [Zn/Fe] ratios at solar metallicity is somewhat more difficult to explain and may require a dilution with essentially zinc-free type SN~Ia ejecta \citep[see][for adiscussion]{DCS17}.

That zinc may be produced through several channels makes it a potentially interesting tracer in the context of Galactic globular cluster (GGCs) evolution. One of the scenarios that was suggested to explain relations between the abundances of light elements observed in GGCs (such as the Na--O, Mg--Al anticorrelation; see, e.g., \citealt{CBG09}) assumes that atmospheres of the second-generation stars could have been polluted with elements synthesized in asymptotic
giant branch (AGB) stars \citep{VD01}. Since zinc may be produced during the $s$-process nucleosynthesis that takes place in intermediate-mass AGB stars (3-6 M$_{\sun}$, e.g., \citealt{KRL09}), relations between its abundance and those of light elements (e.g., sodium) may point to AGB stars as possible polluters. Therefore, a study of the zinc abundance in the atmospheres of GGC stars characterized by different sodium abundances may reveal additional clues about the possible polluters of second-generation stars in the GGCs. As far as individual GGCs are concerned, one potentially interesting target is 47~Tuc, which not only exhibits various correlations and anticorrelations between the abundances of light elements, such as Li, O, Na, Mg, and Al \citep{CBG09,DLG10,DKB14}, but also shows indications for connections between their abundances in the atmospheres of individual cluster stars and kinematical properties of these stars \citep[see, e.g.,][]{KDB14}.

Only a few\ determinations of zinc abundances in 47~Tuc are available, and they are typically based on relatively small samples of stars. For example, \citet[][]{TSA14} determined the zinc-to-iron ratio in 13 RGB stars of this cluster, with a median value of ${\rm[Zn/Fe]}^{\rm 1D~LTE}=0.26\pm0.13$ and no evidence for an intrinsic zinc abundance spread. More recently, based on the study of 19 RGB stars in 47~Tuc, \citet[][]{DCS17} obtained a mean value of $\langle{\rm [Zn/Fe]}\rangle^{\rm 1D~NLTE}=0.17\pm0.10$ (1D NLTE--LTE abundance corrections for zinc are small, < 0.1 dex, see Sect.~\ref{sect:1D_LTE_zinc_abund}; thus, we directly compared LTE and NLTE estimates here). This relatively limited information about the zinc abundance in 47~Tuc is rather unfortunate, especially because at the metallicity of 47~Tuc, $\FeoH^{\rm 1D~LTE}=-0.76$ \citep[][]{CBG09}, \ion{Zn}{I} lines located at 472.2\,nm and 481.0\,nm are quite strong in the spectra of its RGB stars (7--10\,pm) and are only weakly blended in their wings, which makes them good targets for investigating the zinc abundance.

\begin{figure}[t!]
        \begin{center}
                \mbox{\includegraphics[width=7.0cm]{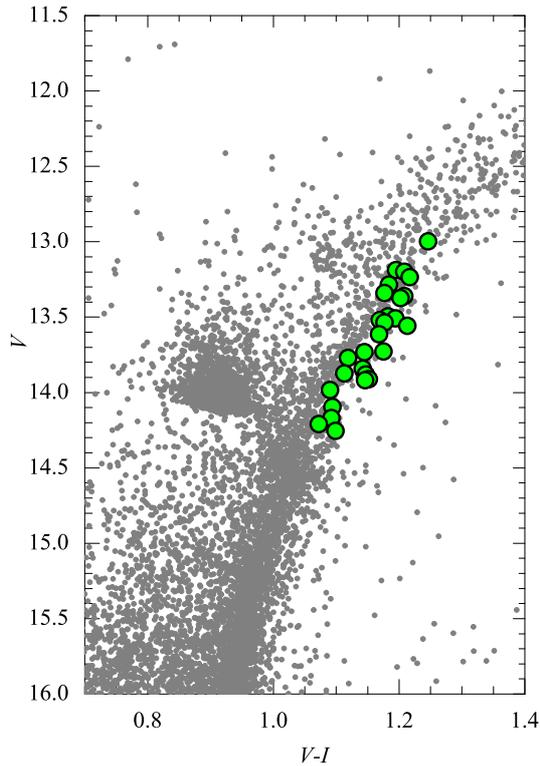}}
                \caption{Upper part of the color-magnitude diagram of 47 Tuc constructed using $VI$ photometry from \citet[][]{BS09}. RGB stars used in this study are marked as filled green circles.\label{fig:cmd}}
        \end{center}
\end{figure}

The main goal of this study was therefore to determine the zinc abundance in a larger sample of RGB stars in 47~Tuc and to search for possible relations between the zinc abundance and those of light elements, primarily sodium and magnesium (but also potassium, see Sect.~\ref{sect:discuss} for a discussion). In addition, we also aimed to study possible connections between the kinematical properties of RGB stars and the zinc abundance in their atmospheres.

The paper is structured as follows: spectroscopic data and the determination of the atmospheric parameters of the target stars are presented in Sect.~\ref{sect:obs_data}, while the method
we used to determine the zinc abundance and its uncertainties is described in Sect.~\ref{sect:1D_LTE_zinc_abund}. The influence of convection on the \ion{Zn}{i} line formation in red giant atmospheres is investigated in Sect~\ref{sect:abn_corr}. The determined zinc abundances, as well as their possible relations with those of light elements and kinematical properties of RGB stars, are discussed in Sect.~\ref{sect:discuss}. Finally, the
main results obtained in this work are summarized in Sect.~\ref{sect:conclus}.

%%%%%%%%%%%%%%%%%%%%%%%%%%%%%%%%%%%%%%%%%%%%%%%%%%%%%%%%%%%%%%%%%%%%%%%%%%%%%%%%%%%%%%%%%
\section{Methodology}\label{sect:method}
%%%%%%%%%%%%%%%%%%%%%%%%%%%%%%%%%%%%%%%%%%%%%%%%%%%%%%%%%%%%%%%%%%%%%%%%%%%%%%%%%%%%%%%%%

The abundance of zinc was determined using 1D hydrostatic \ATLAS\ model atmospheres and a 1D~LTE abundance analysis methodology. A brief description of all steps involved in the abundance analysis is provided below.

%===============================================================================
\subsection{Spectroscopic data and atmospheric parameters of the sample stars\label{sect:obs_data}}
%===============================================================================

We used the same sample of RGB stars of 47~Tuc as in \citet[][hereafter Paper~I]{CKK17}, as well as the same set of archival high-resolution \twodFH\ spectra  obtained with the Anglo-Australian Telescope (AAT)\footnote{The raw spectra were downloaded from the AAT data archive \path{http://site.aao.gov.au/arc-bin/wdb/aat_database/observation_log/make}.}. The spectra were acquired during the science verification phase of the GALAH survey \citep{DFB15} in the wavelength range of $471.5-490.0$\,nm ({\tt HERMES/BLUE}) using a spectral resolution of $R\sim28000$ and exposure time of 1200\,s. The typical signal–to–noise ratio (per pixel) in this wavelength range was $S/N\approx50$. Further details regarding the reduction and continuum normalization of the observed spectra are provided in Paper~I.

The final sample of RGB stars for which zinc abundances were determined contained 27 stars (see Fig.~\ref{fig:cmd}). Their effective temperatures and surface gravities were taken from Paper~I: the former were determined using photometric data from \citet[][]{BS09} and $T_{\rm eff} - (V-I)$ calibration of \citet[][]{RM05}, while the latter were estimated using the classical relation between the surface gravity, mass, effective temperature, and luminosity.

\begin{table}[t!] % liniju parametrai
        \caption{Atomic parameters of \ion{Zn}{i} lines that were used in this study. Natural ($\gamma_{\rm{rad}}$), Stark ($\frac{\gamma_4}{N_{\rm{e}}}$), and van der 
                Waals ($\frac{\gamma_6}{N_{\rm{H}}}$) broadening constants computed using classical 
                prescription are provided in the last three columns.}
        \label{tbl:param}
        \vspace{-5mm}
        \begin{center}
                \scalebox{0.95}{
                        \setlength{\tabcolsep}{2pt}
                        \begin{tabular}{lcccccc}
                                \hline
                                \hline
                                Element & $\lambda$, nm & $\chi$, eV & log$\textit{gf}$ & log$\gamma_{\rm{rad}}$ & log$\frac{\gamma_4}{N_{\rm{e}}}$ & log$\frac{\gamma_6}{N_{\rm{H}}}$ \\
                                \hline
                                \ion{Zn}{I}     &   472.21      &  4.02      &     $-0.33 $     &     7.99           &   -6.21                &         -7.87  \\ 
                                \ion{Zn}{I}     &   481.05      &  4.07      &     $-0.13 $     &     7.98           &   -6.19                &         -7.87  \\
                                \hline
                        \end{tabular}
                        \label{elem}
                }
        \end{center}
\end{table}
%===============================================================================
\subsection{Determination of the 1D~LTE zinc abundance in the atmospheres of RGB stars in 47 Tuc \label{sect:1D_LTE_zinc_abund}}
%===============================================================================

Zinc abundances in individual RGB stars were determined using two \ion{Zn}{i} lines located at 472.21\,nm and 481.05\,nm. Excitation potentials and oscillator strengths of these spectral lines were taken from the VALD-3 database \citep[][]{PKRWJ95,KD11}. Broadening constants were computed using the classical prescription from \citet{C05}. All atomic line parameters are provided in Table~\ref{tbl:param}.

\begin{figure}[t]
        \centering
        \mbox{\includegraphics[width=8.5cm]{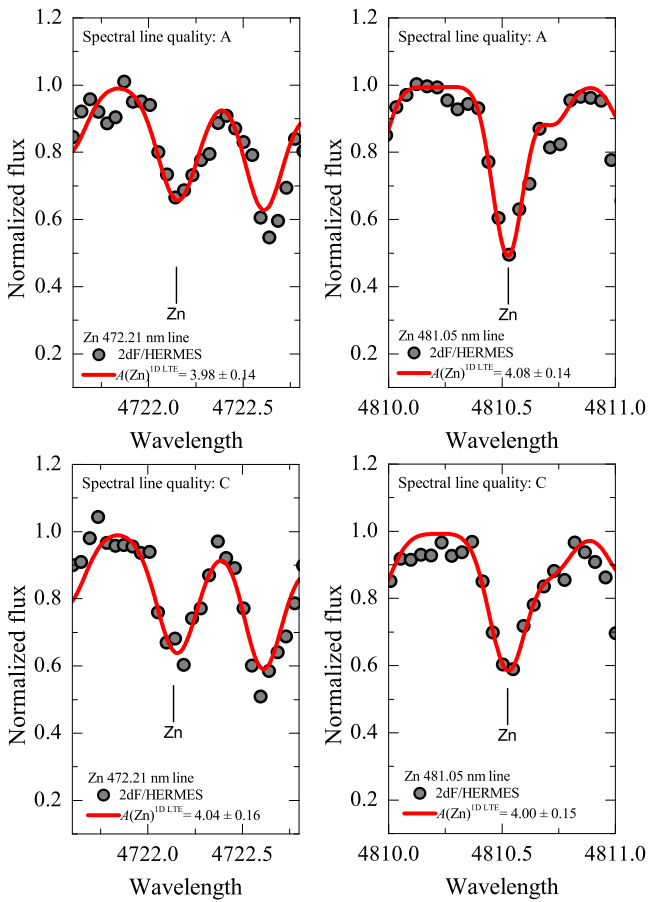}}
        \caption{Typical fits of synthetic spectra (red solid lines) to the observed \ion{Zn}{i} line profiles (marked by vertical ticks) in the {\tt 2dF/HERMES} spectra (filled gray circles) of the target RGB stars S1800 (top row; $\Teff=4510$\,K, $\log g=1.90$)  and S167 (bottom row; $\Teff=4390$\,K, $\log g=1.60$). We also indicate zinc abundances $A{\rm (Zn)}$ determined from each spectral line, together with their errors (Sect.~\ref{sect:1D_LTE_zinc_abund}). The quality class of the \ion{Zn}{i} line is indicated at the top of each panel (see Sect.~\ref{sect:1D_LTE_zinc_abund} for the line class definition).}
        \label{fig:line_profiles}
\end{figure}

The spectra of all RGB stars used in this study were carefully inspected for blends and/or possible contamination of \ion{Zn}{i} lines with telluric lines. We verified that \ion{Zn}{i} lines were not affected by telluric lines, therefore we did not correct the spectra for telluric absorption. Since \ion{Zn}{i} lines were of different quality in the spectra of different stars, the lines were grouped into three classes according to their quality, which was determined by visual inspection:

\begin{list}{--}{}
        \item A-class: strong or moderately strong lines with well-resolved line profiles;
        \item B-class: lines that were moderately blended or insufficiently 
        resolved in the line wings;
        \item C-class: weak, poorly resolved, or significantly blended lines.
\end{list}

\begin{figure}[t]
        \begin{center}
                \mbox{\includegraphics[width=9.0cm]{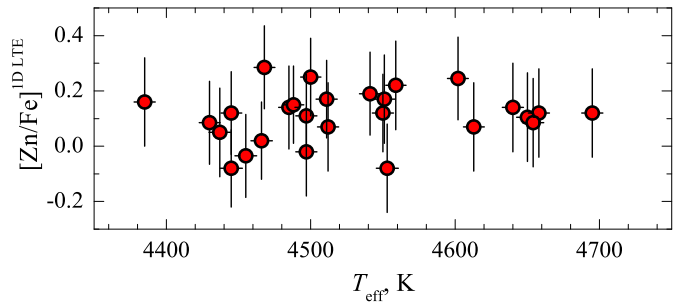}}
                \caption{[Zn/Fe] abundance ratios determined in the sample of 27 RGB stars in 47~Tuc and plotted vs. the effective 
                        temperature of individual stars.\label{fig:ZnTeff}}
        \end{center}
\end{figure}

Zinc abundances were determined using synthetic 1D~LTE spectra that were computed using the \SYNTHE\ package and 1D hydrostatic \ATLAS\ model atmospheres \citep{K93}, with both \SYNTHE\ and \ATLAS\ packages used in the form of the Linux port by \citet{SBC04} and \citet{S05}. Several typical examples of synthetic spectra fits are provided in Fig.~\ref{fig:line_profiles}. As in Paper~I, a fixed value of microturbulence velocity, $\xi=1.5$\,km/s, was used to compute all synthetic spectra. The macroturbulence velocity was varied as a free fitting parameter to obtain the best match to the observed \ion{Zn}{i} lines, with its value changing in the range from 1 to 6 km~s$^{-1}$ in different stars of our sample. We verified that there is no dependence of the determined zinc abundances on the effective temperature (Fig.~\ref{fig:ZnTeff}).

The solar zinc abundance used in our analysis was $\textit{A}(\rm Zn)^{1D~LTE}_{\odot}=4.62 \pm 0.04$. As described in Appendix~\ref{app_sect:Sun}, this value was obtained from the same spectral lines that were used in the analysis of zinc abundance in 47~Tuc (Table~\ref*{tbl:param}). The obtained value agrees well with the previous determinations of the solar zinc abundance, for example, $\textit{A}(\rm Zn)^{1D~LTE}_{\odot}=4.58$ obtained using the same \ion{Zn}{i} lines by \citet{THT05}.

As discussed in Paper~I, we did not determine the iron abundance in individual RGB stars because the number of available \ion{Fe}{i} lines was small and the quality of the spectra did not allow obtaining reliable estimates of the iron abundance. Instead, the fixed value of $\FeoH^{\rm 1D~LTE}=-0.76$ taken from \citet[][]{CBG09} was used for all stars in this study. To check the validity of this assumption, we determined the iron abundance for three stars with the best-quality spectra in our sample, S167, S1490, and S1563 ($\Teff = 4385, 4560, 4695$\,K and $\log g=1.60, 2.05, 2.30$, respectively). In each case, we could identify six \ion{Fe}{i} lines suitable for the abundance analysis. The  iron-to-hydrogen ratios determined in these stars were ${\feoh}^{\rm 1D~LTE}=-0.77, -0.73$, and --0.76, respectively, with the mean value, $\langle{\feoh}^{\rm 1D~LTE}\rangle=-0.75$, nearly identical to $\FeoH^{\rm 1D~LTE}=-0.76$ used in this work. As explained above, however, determination of iron abundances for all stars in our sample was not feasible.

\begin{table}
        \centering
        \caption{The 3D--1D abundance corrections, $\Delta_\mathrm{3D-1D~LTE}$, computed for 
                different strengths of \ion{Zn}{i} lines used in this work (see text for details).}
        \label{tab:abund_corr}
        \centering
        \begin{tabular}{lccccc}
                \hline\hline 
                Element      & $\lambda_{\rm central}$ & \multicolumn{3}{c}{$\Delta_\mathrm{3D-1D~LTE}$, dex} \\
                &    nm                   & weak    & strong                         \\
                \hline
                \ion{Zn}{i}  & 472.21              & $0.039 $  & $0.047 $ \\
                \ion{Zn}{i}  & 481.05              & $0.039 $  & $0.052$  \\
                \hline 
        \end{tabular}
\end{table}     

\begin{figure*}[!t]
\begin{center}
        \mbox{\includegraphics[width=17cm]{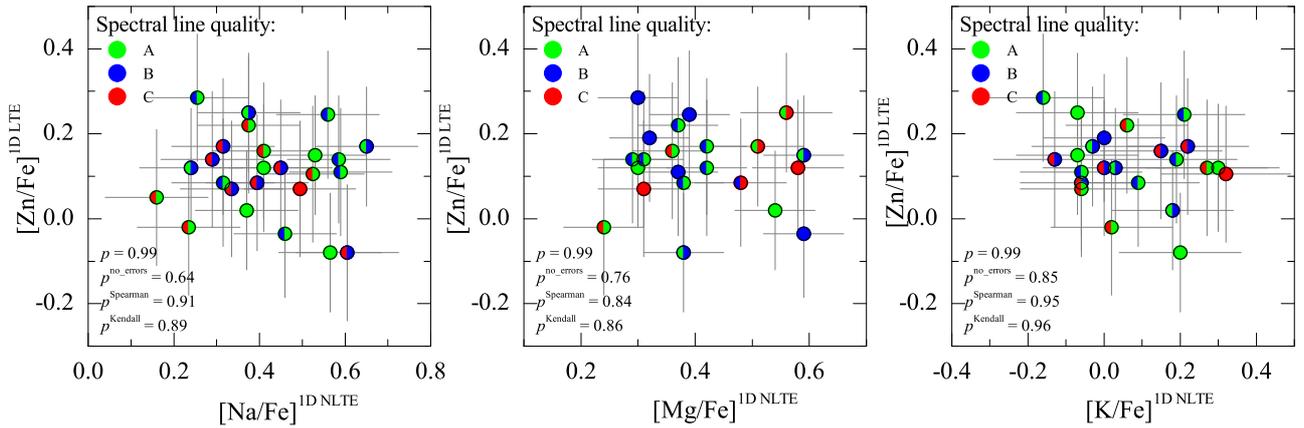}}
        \caption{[Zn/Fe] abundance ratios determined in the sample of 27 RGB stars in 47~Tuc and plotted versus [Na/Fe], [Mg/Fe], and [K/Fe] ratios obtained in the same stars in Paper~I. The
color on the left and right sides of the symbols denote quality class of the spectral lines used to determined abundance ratios on the $y$ and $x$ axes, respectively (see Sect.~\ref{sect:1D_LTE_zinc_abund}). We also list the values of two-tailed probabilities, $p$, computed using Pearson’s parametric correlation coefficients (with and without abundance errors, $p$ and $p^{\rm no\_errors}$, respectively), as well as Spearman’s ($p^{\rm Spearman}$) and Kendall’s ($p^{\rm Kendall}$) non-parametric rank-order correlation coefficients (see Sect.~\ref{sect:discuss}).
        \label{fig:abn-ratios}}
\end{center}
\end{figure*}

We did not correct the obtained zinc abundances for NLTE effects. To our knowledge, the only available source of published 1D~NLTE--LTE abundance corrections for the two \ion{Zn}{i} lines used in our study is \citet{THT05}. However, the authors do not provide corrections at the effective temperatures and gravities bracketing those of our sample stars because of numerical problems in their NLTE computations at these atmospheric parameters. 
Nevertheless, by extrapolating their available data, we estimated that abundance corrections for our sample stars will be on the order of $\Delta_{\rm 1D~NLTE-LTE}\approx -0.05$\,dex and are
therefore of little significance in the context of present study.

Uncertainties in the determined zinc abundances were evaluated using the same methodology as in Paper~I, which is described in Appendix~\ref{app_sect:abund_err}. Our basic assumption was that these uncertainties were governed by the errors in the determined stellar atmospheric parameters and uncertainties in the line profile fitting. We stress that we did not take into account errors due to uncertainties in the atomic line parameters. Therefore, the obtained error estimates (Table~\ref{tab_app:abund_err}) only provide a lower limit for the uncertainties in the zinc abundances determined from individual \ion{Zn}{i} lines because they do not account for other possible sources of systematic errors. Moreover, the individual errors are considered to be uncorrelated.

%===============================================================================
\subsection{3D--1D~LTE zinc abundance corrections\label{sect:abn_corr}}
%===============================================================================

The influence of 3D hydrodynamical effects on the formation of \ion{Zn}{i} lines has so far been investigated only for dwarf and subgiant stars \citep[][]{NCA04}. The authors have found that the role of these effects was very small, with the resulting 3D--1D~LTE abundance corrections significantly below 0.1\,dex. 

To investigate the 3D effects of the formation of \ion{Zn}{i} lines in the atmospheres of RGB stars in 47~Tuc, we used 3D hydrodynamical \COBOLD\ \citep[][]{FSL12} and 1D hydrostatic \LHD\ \citep{CLS08} model atmospheres. For this, we used models with atmospheric parameters similar to those of the median object in our RGB star sample, $\Teff\approx4490$\,K,  $\log g= 2.0$, and $\moh=-1.0$. This choice is justified because the range covered by the atmospheric parameters of the sample stars is small ($\Delta \Teff\approx300$\,K, $\Delta \log g\approx0.3$), which means that the obtained corrections should be applicable to all RGB stars in our sample. In addition
to the atmospheric parameters, the \COBOLD\ and \LHD\ models share identical chemical composition, opacities, and equation of state (see Paper~I for details). Spectral line synthesis computations were carried out with the \LINFOR\ spectral synthesis package\footnote{\url{http://www.aip.de/Members/msteffen/linfor3d}}.

The 3D--1D~LTE abundance corrections, $\Delta_{\rm 3D-1D~LTE}$, were computed for two values of the line equivalent width, $W$ (corresponding to ``weak'' and ``strong'' spectral lines) that bracketed the range measured in the observed spectra of the sample RGB stars. Equivalent widths used for the weakest lines were 8.5\,pm and 8\,pm, while for the strongest lines we used 10.5\,pm and 10\,pm for Zn 472.21 and Zn 481.05, respectively. The microturbulence velocity in the 3D model atmosphere was determined by applying  Method~1 \footnote{The microturbulence velocity shows very little variation (<~0.01 km/s) within the range of line strengths \st{we} used to calculate the 3D--1D abundance corrections.} described in \citet[][]{SCL13} and was subsequently used in the spectral line synthesis with the \LHD\ model atmospheres (see Paper~I). 

The obtained 3D--1D~LTE abundance corrections are provided in Table~\ref{tab:abund_corr}. For both lines, they do not exceed $0.05$\,dex and show little dependence on the line strength, which allows us to conclude that the influence of convection on the formation of \ion{Zn}{i} lines in the atmospheres of RGB stars is minor. The value of the 3D--1D~LTE abundance correction for \ion{Zn}{i} line at 481.05 nm is indeed very similar to that obtained using the same \COBOLD\ and \LHD\ models by \citet{DCS17}. 

We note that the computed abundance corrections were not used to obtain the 3D-corrected abundances by adding the 3D--1D~LTE corrections to the determined 1D~LTE abundances of zinc. The main reason for this was that the obtained $\Delta_{\rm 3D-1D~LTE}$ corrections were small. Therefore, if applied, they result in a very small and nearly uniform shift of abundances determined in all RGB stars, with no effect on the intrinsic abundance spread of zinc and/or various possible relations in the [Zn/Fe] -- [Na/Fe] and [Zn/Fe] -- [Mg/Fe] planes.

%%%%%%%%%%%%%%%%%%%%%%%%%%%%%%%%%%%%%%%%%%%%%%%%%%%%%%%%%%%%%%%%%%%%%%%%%%%%%%%%%%%%%%%%%
\section{Results and discussion \label{sect:discuss}}
%%%%%%%%%%%%%%%%%%%%%%%%%%%%%%%%%%%%%%%%%%%%%%%%%%%%%%%%%%%%%%%%%%%%%%%%%%%%%%%%%%%%%%%%%

The mean value of the zinc-to-iron abundance ratio we obtained in the sample of 27 RGB stars is $\langle{\rm[Zn/Fe]}\rangle^{\rm 1D~LTE} = 0.11\pm0.09$, where the number after the $\pm$ sign is the RMS abundance variation due to the star-to-star scatter. Since this variation is not insignificant, the question arises whether it might be a result of intrinsic scatter in the zinc abundance. To answer this question, we applied the  maximum-likelihood (ML) technique to evaluate the mean zinc-to-iron abundance ratio, $\langle [\rm Zn/Fe]\rangle$, as well as its intrinsic spread, $\sigma_{\rm int}^{[\rm Zn/Fe]}$, in our sample of RGB stars. For this, we followed the procedure used in Paper~I, which was based on the prescription of \citet{MBI12,MBM15}. The  average zinc-to-iron ratio obtained in the ML test is $\langle [\rm Zn/Fe]\rangle = 0.11$ and its uncertainty is $\pm0.03$. We note that the latter value is the error of the mean and therefore is smaller than the RMS variation due to the star-to-star abundance scatter. The determined intrinsic abundance variation is $\sigma_{\rm int}^{[\ion{Zn}{}/\ion{Fe}]}=0.00\pm0.04$, thus there is no intrinsic variation intrinsic variation in the zinc abundance in the RGB stars of 47~Tuc. We stress, however, that this non-detection may be at least partly due to significant uncertainties of the zinc abundance measurements in individual stars, which in turn were caused by the relatively low $S/N$  in the observed spectra.

\begin{figure}[t!]
        \begin{center}
                \mbox{\includegraphics[width=8.3cm]{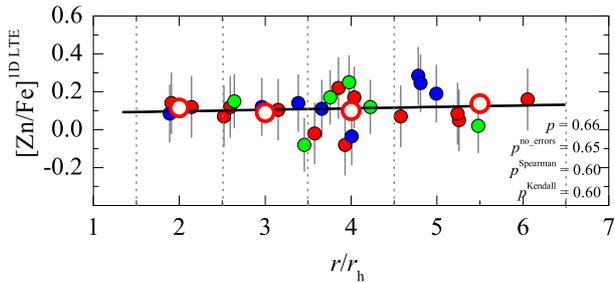}}
                \caption{[Zn/Fe]$^{\rm 1D~LTE}$ abundance ratios determined in the sample of 27 RGB stars in 47~Tuc and plotted vs. the projected radial distance of individual stars (small filled symbols). The symbol colors are identical to those shown in Fig~\ref{fig:abn-ratios}. Large open circles are averages obtained in non-overlapping $\Delta r/r_h = 1$ bins.\label{fig:Zn-dist}}
        \end{center}
\end{figure}

To investigate possible relations between the abundance of zinc and those of other light elements, we used the abundances of Na, Mg, and K determined in our sample RGB stars in Paper~I (all abundances of light elements were determined using the 1D~NLTE methodology). It is well established that the sodium abundance shows a significant intrinsic star-to-star variation and is anticorrelated with that of oxygen in the vast majority of GGCs studied so far. Therefore, the relation between abundances of Na and Zn would indicate that the two elements were synthesized by the same polluters. In some GGCs, the Mg--Al correlation is observed and points to the Mg--Al cycle, which occurs at higher temperatures than the synthesis of Na. Thus, in a similar way, the relation between Zn and Mg abundances may indicate that the nucleosynthesis of the two elements is connected. The case with potassium is less clear. Recently, \citet{MMB17} claimed a detection of a K--Na correlation and K--O anticorrelation in a sample of 144 RGB stars in 47~Tuc. However, our own study of the K abundance in RGB and TO stars in 47~Tuc  (32 and 75 objects, respectively) does not support this claim (Paper~I and \citet[][]{CKK18}, hereafter Paper~II; we note that our data may suggest a weak K--Na anticorrelation in TO stars, but its statistical significance is low). It is worthwhile noting that, if confirmed, relations between the abundance of K and other light elements would be very difficult to explain from the theoretical point of view because K is synthesized at temperatures $\approx2\times 10^8$\,K, at which all sodium should be destroyed \citep[e.g.,][]{PCI17}. Despite this, we wished to verify whether  a relation between the abundance of zinc and potassium in the sample of our RGB stars existed, which in case of a positive detection would add another piece of information that would need to be explained by the chemical evolution models of the GGCs. The yields from AGBs depend on the stellar mass, thus in order to reproduce general trends of the light chemical elements (O, Na, Mg, and Al) seen in different globular clusters, the mass range of AGB stars should be in the range of 5-9 M$_\sun$ \citep{DDV10,VDD12}. Zinc may be synthesized in slightly less massive AGB stars, 3-6 M$_\sun$ \citep{KRL09}, thus there may be a slight overlap in stellar masses in which both zinc and light elements could synthesized in AGB stars. In such a situation, relations between the abundances of zinc and those of light elements would be expected.

\begin{figure}[t!]
        \begin{center}
                \mbox{\includegraphics[width=8.7cm]{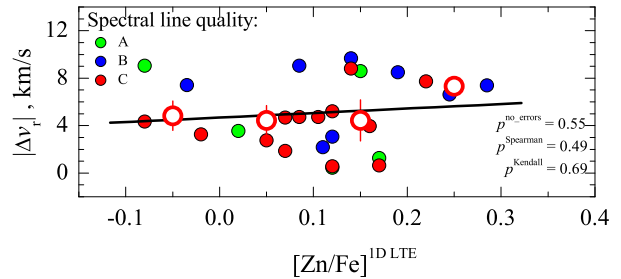}}
                \caption{Absolute radial velocities determined for the sample of 27 RGB stars in 47~Tuc and plotted vs. [Zn/Fe]$^{\rm 1D~LTE}$ ratios of individual stars (small filled symbols). The
symbol colors are identical to those shown in Fig~\ref{fig:abn-ratios}. Large open circles are averages obtained in non-overlapping $\Delta {\rm [Zn/Fe]^{\rm 1D~LTE}} = 0.1$ bins.} \label{fig:Zn-disp}
        \end{center}
\end{figure}

\begin{figure}[t!]
\begin{center}
                \mbox{\includegraphics[width=8.0cm]{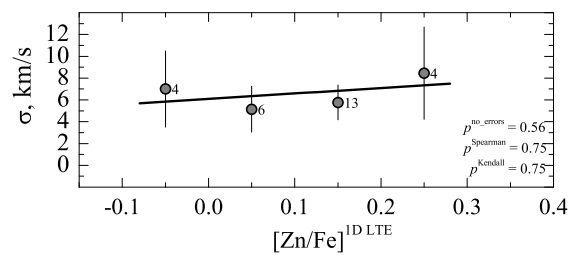}}
                \caption{Radial velocity dispersion of RGB stars in 47~Tuc plotted versus the [Zn/Fe]$^{\rm 1D~LTE}$ ratio (both quantities were computed in non-overlapping $\Delta {\rm [Zn/Fe]^{\rm 1D~LTE}} = 0.1$ bins). The black solid line shows a linear fit to the data.\label{fig:Zn_sigma}}
\end{center}
\end{figure}

The determined [Zn/Fe] ratios are plotted versus various light element element-to-iron ratios in Fig.~\ref{fig:abn-ratios}. Similar to Paper~I, we used Student's $t$-test to verify the validity of the null hypothesis, meaning that there is no correlation in the different panels of Fig.~\ref{fig:abn-ratios}. For this, using each $x-y$ dataset shown in the three panels of Fig.~\ref{fig:abn-ratios}, we computed the two-tailed probability, $p$, that the $t$-value in the given dataset could attain the value we obtained when there is no correlation in the given $x-y$ plane. In all panels of Fig.~\ref{fig:abn-ratios},  Pearson’s correlation coefficients were computed by taking errors on both $x$ and $y$ axes into account (they are marked as $p$ in Fig.~\ref{fig:abn-ratios}). In addition, we also carried out non-parametric Spearman’s and Kendall’s $\tau$ rank-correlation tests, in this case without accounting for errors in the determined abundances. Finally, we also computed $p$-values using Pearson’s correlation coefficients obtained without taking abundance errors into account (marked as $p^{\rm no\_errors}$ in Fig.~\ref{fig:abn-ratios}). These tests were also performed using the data shown in Fig.~\ref{fig:Zn-dist}--\ref{fig:Zn_sigma} (see discussion below). The $p$-values determined in all tests are provided in the corresponding panels of Fig.~\ref{fig:abn-ratios}--\ref{fig:Zn_sigma}. 

The results of these tests do not show any indications of statistically significant relations between the abundances of zinc and the
other light elements shown in Fig.~\ref{fig:abn-ratios}: in all cases, the $p$-values are $\geq0.64$. Similarly, there is no statistically significant relation between the determined [Zn/Fe] ratios and the normalized distance from the cluster center, \textit{r}/\textit{r}$\rm_{h}$ (Fig.~\ref{fig:Zn-dist}): all $p$-values are $\geq 0.60$ (here, \textit{r} is the projected distance from the cluster center and \textit{r}$\rm_{h}$ is the half-light radius of 47~Tuc taken from \citealt[][]{TDK93}, $r_{\rm h}=174^{\prime\prime}$). Neither
do we find a statistically significant relation between the [Zn/Fe] ratios and absolute radial velocities of RGB stars, $\left|\Delta v_{\rm r}\right| \equiv \left| v_{\rm rad} - {\langle v_{\rm rad} \rangle}^{\rm clust} \right|$, where $v_{\rm rad}$ is radial velocity of the individual star and ${\langle v_{\rm rad} \rangle}^{\rm clust}= -18.6$\,km/s is the mean radial velocity of the sample (Fig.~\ref{fig:Zn-disp}). In this test, the absolute radial velocities were taken from Paper~I, while the obtained $p$-values are all $\geq0.49$. Finally, we tested whether there may be a relation between the radial velocity dispersion of individual stars (computed in non-overlapping 0.1\,dex wide bins of [Zn/Fe] ratios) and their average [Zn/Fe] values (Fig.\ref{fig:Zn_sigma}). Again, we find no statistically significant relation, with $p$-values for all tests $\geq0.33$. 

We also performed all statistical tests mentioned above using the three sub-samples of zinc abundances sorted according to the quality class of Zn I lines from which the abundances were determined. For none of the three subsamples (A-C, marked with different symbols in Figs.\ref{fig:abn-ratios}-\ref{fig:Zn-disp}) did we find statistically significant relations in any of the planes shown in Figs.\ref{fig:abn-ratios}-\ref{fig:Zn-disp}.

Our results therefore indicate that there is no relation between the nucleosynthesis of zinc and that of light elements in 47~Tuc. The sample-averaged zinc-to-iron abundance ratio determined in this study, $\langle{\rm[Zn/Fe]}\rangle^{\rm 1D~LTE} = 0.11\pm0.09$, agrees to within less than one sigma to those obtained by \citet[][]{TSA14}, ${\rm[Zn/Fe]}^{\rm 1D~LTE}=0.26\pm0.13$ (13 RGB stars), and \citet[][]{DCS17}, $\langle{\rm [Zn/Fe]}\rangle^{\rm 1D~NLTE}=0.17\pm0.10$ (19 RGB stars). Our value is nearly identical to the average [Zn/Fe] ratio typical to Galactic field stars at the metallicity of 47~Tuc \citep[e.g.,][]{MKS02,BFd15,DCS17,dSBF18}. It also agrees well with the [Zn/Fe] ratio determined at this galactocentric distance for the Galactic red giants of similar metallicity \citep{DCS17,EBA18}. All this evidence suggests that the production of zinc in 47~Tuc most likely followed the same pathways as in Galactic field stars of the same metallicity, and it most likely occurred through $\alpha$-element nucleosynthesis. We find no evidence that zinc could have been synthesized in 47~Tuc during the $s$-process nucleosynthesis.

%%%%%%%%%%%%%%%%%%%%%%%%%%%%%%%%%%%%%%%%%%%%%%%%%%%%%%%%%%%%%%%%%%%%%%%%%%%%%%%%%%%%%%%%%
\section{Conclusions}\label{sect:conclus}
%%%%%%%%%%%%%%%%%%%%%%%%%%%%%%%%%%%%%%%%%%%%%%%%%%%%%%%%%%%%%%%%%%%%%%%%%%%%%%%%%%%%%%%%%

We determined the abundance of zinc in the atmospheres of 27 RGB stars in the Galactic globular cluster 47~Tuc. The abundances were obtained using archival \twodFH\ spectra ($471.5-490.0$\,nm, $\textit{R}=28\,000$, $S/N\approx50$) that were obtained with the Anglo-Australian Telescope. Spectroscopic data were analyzed using 1D \ATLAS\ model atmospheres and the 1D~LTE abundance analysis methodology. The 1D~LTE spectral line synthesis was performed with the \SYNTHE\ package. The obtained sample-averaged zinc-to-iron abundance ratio is $\langle{\rm [Zn/Fe]}\rangle^{\rm 1D~LTE}=0.11\pm0.09$, here the value  following the $\pm$ sign is RMS abundance variation due to star-to-star scatter. We also used 3D hydrodynamical \COBOLD\ and 1D hydrostatic \LHD\ model atmospheres to compute the 3D--1D~LTE abundance corrections for \ion{Zn}{i} lines. The obtained 3D--1D LTE abundance corrections are always positive and are in the range 0.04 to 0.05\,dex, indicating that the influence of convection on the formation of \ion{Zn}{i}  lines in the atmospheres of RGB stars in 47~Tuc is minor. Applying these corrections would lead to a 3D-corrected estimate of the average zinc-to-iron abundance ratio of $\langle{\rm [Zn/Fe]}\rangle^{\rm 1D~LTE}=0.16\pm0.09$.

The averaged $\langle{\rm [Zn/Fe]}\rangle^{\rm 1D~LTE}=0.11\pm0.09$ ratio derived in this study agrees well with the mean [Zn/Fe] ratios in 47~Tuc determined in the previous studies that were
based on smaller samples of stars, ${\rm[Zn/Fe]}^{\rm 1D~LTE}=0.26\pm0.13$ \citep[][13 RGB stars]{TSA14} and $\langle{\rm [Zn/Fe]}\rangle^{\rm 1D~NLTE}=0.17\pm0.10$, \citet[][19 RGB stars]{DCS17}. Our results show no statistically significant relations between the [Zn/Fe]$^{\rm 1D~LTE}$ ratio and abundances of light elements (Na, Mg, K), as well as between the [Zn/Fe]$^{\rm 1D~LTE}$ ratio and kinematic properties of the RGB stars. The sample-averaged [Zn/Fe]$^{\rm 1D~LTE}$ ratio coincides with the mean [Zn/Fe] ratio obtained in Galactic field stars at the metallicity of 47~Tuc. All these facts indicate that the nucleosynthesis of zinc and that of light elements, such as Na and Mg, has proceeded separately in 47~Tuc. The obtained average [Zn/Fe]$^{\rm 1D~LTE}$ ratio suggests that in 47~Tuc zinc is slightly enhanced at this metallicity and most likely originated during the $\alpha$-element nucleosynthesis.

%===============================================================================
\begin{acknowledgements}
%===============================================================================

Based on data acquired through the Australian Astronomical Observatory, under program 2013B/13. This work was supported by grants from the Research Council of Lithuania (MIP-089/2015, TAP LZ 06/2013). H.G.L. acknowledges financial
support by the Sonderforschungsbereich SFB\,881 ``The Milky Way System''
(subprojects A4) of the German Research Foundation (DFG).

\end{acknowledgements}

%===============================================================================
\bibliographystyle{aa}
%===============================================================================

\
%===============================================================================

\begin{appendix}

%%%%%%%%%%%%%%%%%%%%%%%%%%%%%%%%%%%%%%%%%%%%%%%%%%%%%%%%%%%%%%%%%%%%%%%%%%%%%%%%%%%%%%%%%
\section{Zinc abundance in the Sun \label{app_sect:Sun}}
%%%%%%%%%%%%%%%%%%%%%%%%%%%%%%%%%%%%%%%%%%%%%%%%%%%%%%%%%%%%%%%%%%%%%%%%%%%%%%%%%%%%%%%%%

We determined the solar zinc abundance using the same \ion{Zn}{i} lines and their atomic parameters (Table~\ref{tbl:param}) as used by us in the analysis of the zinc abundance in RGB stars in 47~Tuc. Synthetic \ion{Zn}{i} line profiles were computed using the \SYNTHE\ spectral synthesis package and 1D hydrostatic \ATLAS\ model atmospheres (see Sect.~\ref{sect:1D_LTE_zinc_abund}). Atomic line parameters were taken from the VALD-3 atomic database \citep[][]{PKRWJ95,KD11}.

\begin{figure}[tb]
        \begin{center}
                \mbox{\includegraphics[width=9cm]{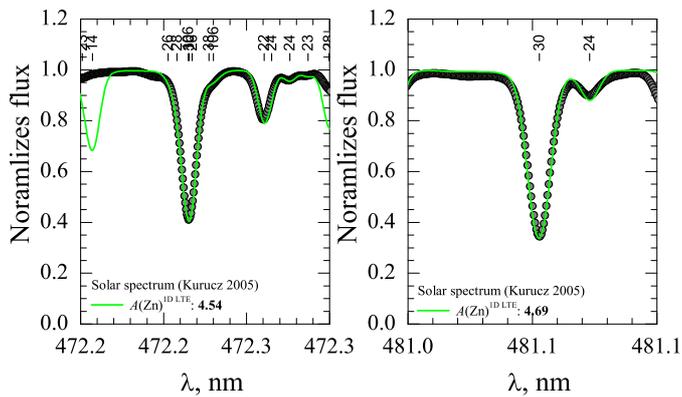}}
                \caption{\ion{Zn}{i} lines (filled gray circles) observed at 472.2 nm and 481.0\,nm in the solar spectrum. The synthetic solar spectrum is shown as a green solid line. Numbers above the spectrum correspond to the atomic numbers of the line-producing chemical element.\label{fig:Zn-Sun}}
        \end{center}
\end{figure}

To determine the solar zinc abundance, we used the re-reduced Kitt Peak Solar Flux atlas of \citet{K06}. The solar spectrum covers a wavelength range of 300--1000\,nm, with $\textit{R}=523\,000$, and $S/N\approx 3500$ in the wavelength range around \ion{Zn}{i} lines. The solar 1D~LTE zinc abundance was determined using the spectral lines located at 472.2 nm and 481.0 nm, with the microturbulence velocity set to $1.0$\,km/s. Each line was fit independently with the synthetic spectrum (Fig.~\ref{fig:Zn-Sun}) that was computed using an \ATLAS\ model atmosphere of the Sun ($\Teff = 5777$\,K and $\logg=4.43$, as recommended by \citealt[][]{ASD16}). The solar abundances of zinc obtained from the 472.2 nm and 481.0\,nm \ion{Zn}{i} lines were $\textit{A}(\rm Zn)^{1D~LTE}_{\odot}=4.54$ and $\textit{A}(\rm Zn)^{1D~LTE}_{\odot}=4.69$, respectively, with a mean value of $\textit{A}(\rm Zn)^{1D~LTE}_{\odot}=4.62\pm0.08$ (the value after the ``$\pm$" sign is the RMS variation). This agrees well with those determined in the earlier studies, for
example, $\textit{A}(\rm Zn)^{1D~LTE}_{\odot}=4.58$ obtained using the same \ion{Zn}{i} lines by \citet{THT05}, $\textit{A}(\rm Zn)^{1D~LTE}_{\odot}=4.62$ derived by \citet[][]{LPG09}, and $\textit{A}(\rm Zn)^{1D~LTE}_{\odot}=4.56\pm0.05$ recommended by \citet{AGS09}.

\begin{table}[t!]
        \begin{center}
                \caption{Total error in the abundance of zinc determined from the \ion{Zn}{i} lines in the atmosphere of the Sun, $\sigma(A)_{\rm tot}$,. The sign $\pm$ or $\mp$ reflects the change in the elemental abundance that occurs due to the increase (top sign) or decrease (bottom sign) in the uncertainty from a given error source (as indicated in column header; see Paper~II for details).}
                \vspace{-2mm}
                \resizebox{\columnwidth}{!}{            
                        \begin{tabular}{lccccccccc}
                                \hline\hline
                                Element & $\lambda$ & $\sigma(\Teff)$ & $\sigma(\log g)$  &      $\sigma(\xi_{\rm t})$    &  $\sigma(\rm cont)$ &  $\sigma(\rm fit)$  & $\sigma(A)_{\rm tot}$  \\ [0.5ex]
                                & nm          &       $\pm10$\,K,              &             $\pm0.02$           &     $\pm0.06$\,km/s         &          &          &     \\
                                \hline
                                \textbf{Sun}     & & & & & & & &                                                                            \\
                                \ion{Zn}{i}      & 472.21 & $\pm 0.005$ & $\mp 0.001$  & $\mp 0.023$ & $\pm 0.015$  & $\pm 0.017$ &  0.032  \\
                                \ion{Zn}{i}      & 481.05 & $\pm 0.011$ & $\mp 0.005$  & $\mp 0.032$ & $\pm 0.016$  & $\pm 0.019$ &  0.042  \\
                                \hline
                        \end{tabular}}
                        \label{tab:abnd_err_Sun}
                \end{center}
                \vspace{-5mm}
        \end{table}

\begin{table}[t!]
        \begin{center}
                \caption{Errors in the abundance of zinc determined from the two \ion{Zn}{i} lines used in this study and different spectral line quality (A--C, see Sect.~\ref{sect:discuss}). The $\pm$ or $\mp$ sign reflects the change in elemental abundance that occurs due to increase (top sign) or decrease (bottom sign) in a given atmospheric parameter. For example, an increase in the effective temperature leads to an increase in the abundance of zinc ($\pm$), while increasing microturbulence velocity results in a decreasing zinc abundance ($\mp$).}
                \vspace{-5mm}
                \resizebox{\columnwidth}{!}{            
                        \begin{tabular}{lccccccccc}
                                \hline\hline
                                Element & Line           & Line             &  $\sigma(\Teff)$  & $\sigma(\log g)$  &      $\sigma(\xi_{\rm t})$    &  $\sigma(\rm cont)$ &  $\sigma(\rm fit)$  & $\sigma(A)_{\rm tot}$  \\ [0.5ex]
                                & $\lambda$, nm  & quality          &      dex          &             dex   &        dex                    &   dex               &    dex              &   dex                  \\
                                \hline
                                \ion{Zn}{I}   & 472.21 & A & $\pm0.03$ & $\mp0.04$  & $\mp0.11$ & $0.04$  & $0.06$ &  $0.14$ \\
                                &                      & B & $\pm0.03$ & $\mp0.04$  & $\mp0.11$ & $0.04$  & $0.08$ &  $0.15$ \\
                                &                      & C & $\pm0.03$ & $\mp0.04$  & $\mp0.11$ & $0.04$  & $0.09$ &  $0.16$ \\
                                \ion{Zn}{I}   & 481.05 & A & $\pm0.03$ & $\mp0.03$  & $\mp0.11$ & $0.05$  & $0.05$ &  $0.14$ \\
                                &                      & B & $\pm0.03$ & $\mp0.03$  & $\mp0.11$ & $0.05$  & $0.06$ &  $0.14$ \\
                                &                      & C & $\pm0.03$ & $\mp0.03$  & $\mp0.11$ & $0.05$  & $0.08$ &  $0.15$ \\
                                \hline
                        \end{tabular}}
                        \label{tab_app:abund_err}
                \end{center}
                
                \vspace{-5mm}
        \end{table}

We further estimated the uncertainty of the solar zinc abundance determination using the same procedure and uncertainties in the atmospheric parameters of the Sun and spectral line fitting procedure as described in Paper~II. Errors from the individual uncertainties are provided in Table~\ref{tab:abnd_err_Sun}. The final value of the solar zinc abundance obtained from the two \ion{Zn}{i} lines is $\textit{A}(\rm Zn)^{1D~LTE}_{\odot}=4.62 \pm 0.04$.

%%%%%%%%%%%%%%%%%%%%%%%%%%%%%%%%%%%%%%%%%%%%%%%%%%%%%%%%%%%%%%%%%%%%%%%%%%%%%%%%%%%%%%%%%
\section{Uncertainties in zinc abundances determined in the RGB stars in 47~Tuc}\label{app_sect:abund_err}
%%%%%%%%%%%%%%%%%%%%%%%%%%%%%%%%%%%%%%%%%%%%%%%%%%%%%%%%%%%%%%%%%%%%%%%%%%%%%%%%%%%%%%%%%
        
The uncertainty in the determined zinc abundances was estimated following the same procedure as in Paper~I. For this, we assumed that the error in the determined abundances occurs due to an
inaccurate determination of the atmospheric parameters ($\Teff$, $\log g$, $\xi_{\rm t}$) and the spectral line profile fitting, with the additional assumption that the latter depends on the goodness of line profile fit and the placement of continuum level. These uncertainties were estimated in the following way:

\begin{list}{--}{}
        
        \item \emph{\textup{Errors in the determination of atmospheric parameters}}:
        Following Paper~I, we assumed that the error in the determination of the effective temperature of the RGB stars was $\pm65$\,K. 
        The errors in the determined zinc abundance resulting from this uncertainty, $\sigma(\Teff)$, are provided in Table~\ref{tab_app:abund_err}, Col.~4, for both \ion{Zn}{i} lines. For the uncertainty in surface gravity, we adopted a conservative (and, in our view, more realistic) value $\pm0.1$\,dex; the resulting errors in the zinc abundances, $\sigma(\log g)$, are listed in Table~\ref{tab_app:abund_err}, Col.~5. Since we did not derive individual values of the microturbulence velocity for individual RGB stars, we adopted an uncertainty in the  microturbulence velocity of $\pm 0.15$\,km/s. This value is based on the RMS variation of microturbulence velocity determined in RGB stars in 47~Tuc by \citet[][58 objects]{CBG09} and \citet[][81 objects]{CPJ14}, with stars occupying the same range in the effective temperature as those used in our sample. The obtained abundance uncertainties that occur due to errors in microturbulence velocity, $\sigma(\xi_{\rm t})$, are listed in Table~\ref{tab_app:abund_err}, Col.~6.
        
        \item \emph{\textup{Errors in the spectral line profile fitting}}: The error in continuum determination was estimated in the same way as in Paper~I, that is, by measuring the dispersion at the continuum level (inverse signal-to-noise ratio) in the spectral regions deemed to be free of spectral lines (see Appendix~B in Paper~I for details). The errors in the line profile fitting were computed individually for each line quality class (see Sect. \ref{sect:1D_LTE_zinc_abund}) using RMS values as a measure of differences between the observed and best-fit synthetic line profiles. These RMS values were used to compute the change in the determined abundance of zinc. The final errors due to continuum determination and spectral line profile fitting are shown in Table~\ref{tab_app:abund_err}, Cols.~7 and 8, respectively.  
        
\end{list}

The errors in the determined zinc abundances resulting from various uncertainty sources are listed in Table~\ref{tab_app:abund_err}. The total uncertainties in zinc abundance (obtained by adding individual components in quadratures, Col.~9 in Table~\ref{tab_app:abund_err}) are shown in Fig.\ref{fig:ZnTeff}, \ref{fig:abn-ratios}, and \ref{fig:Zn-dist}. These errors were also used in the ML analysis aimed to determine the average value of the zinc abundance and its intrinsic spread in the sample of RGB stars (see Sect.~\ref{sect:discuss}). 

We stress that in this procedure, errors resulting from uncertainties in the atomic line parameters were ignored. Therefore, the total abundance uncertainties listed in Table~\ref{tab_app:abund_err} are only lower limit estimates because they do not account for the various systematic uncertainties that are unavoidable in the abundance analysis procedure.

%%%%%%%%%%%%%%%%%%%%%%%%%%%%%%%%%%%%%%%%%%%%%%%%%%%%%%%%%%%%%%%%%%%%%%%%%%%%%%%%%%%%%%%%%
%%%%%%%%%%%%%%%%%%%%%%%%%%%%%%%%%%%%%%%%%%%%%%%%%%%%%%%%%%%%%%%%%%%%%%%%%%%%%%%%%%%%%%%%%

\end{appendix}
\end{document}